\documentclass[aps,prd,twocolumn,showpacs,amsmath,amssymb,nofootinbib,superscriptaddress]{revtex4-2}
\usepackage{graphicx}  
\usepackage{color}
\usepackage{times}
\usepackage[utf8]{inputenc} 
\usepackage{float}
\usepackage{bm}
\usepackage{ulem}
\usepackage{multirow}
\usepackage{makecell}
\usepackage{url}
\usepackage{natbib}
\usepackage{mathrsfs}
\usepackage{physics}
\usepackage{comment}
\usepackage[table,xcdraw]{xcolor}
\usepackage{booktabs,makecell}
\usepackage{amsmath}
\usepackage{pifont}
\usepackage[colorlinks=true,citecolor=blue,urlcolor=blue,linkcolor=blue]{hyperref}
\usepackage{microtype}
\usepackage[none]{hyphenat}

\newcommand{\dalm}{\kern1pt\vbox{\hrule height 0.9pt\hbox{\vrule width 0.9pt
\hskip 2.5pt\vbox{\vskip 5.5pt}\hskip 3pt\vrule width 0.3pt}\hrule height 0.3pt}
\kern1pt}

\begin{document}



\title{Emergence of new oscillation modes in dark matter admixed neutron stars}


\author{Hajime Sotani}
\email{sotani@yukawa.kyoto-u.ac.jp}
\affiliation{Department of Mathematics and Physics, Kochi University, Kochi, 780-8520, Japan}
\affiliation{RIKEN Center for Interdisciplinary Theoretical and Mathematical Sciences (iTHEMS), RIKEN, Wako 351-0198, Japan}
\affiliation{Theoretical Astrophysics, IAAT, University of T\"{u}bingen, 72076 T\"{u}bingen, Germany}

\author{Ankit Kumar}
\email{ankit.k@iopb.res.in}
\affiliation{Department of Mathematics and Physics, Kochi University, Kochi, 780-8520, Japan}

\date{\today}

\begin{abstract}
Dark matter admixed neutron stars provide a promising avenue for observationally probing the dark matter characteristic. In this study, we examine non-radial oscillations in neutron stars containing self-interacting dark matter, which interacts with normal matter exclusively via gravity. To achieve this, we derive a new set of perturbation equations for a multi-fluid system under the Cowling approximation. Using these equations, we analyze the oscillation spectra and identify additional modes associated with dark matter, alongside those of normal matter. We find that the frequency behavior becomes more intricate with increasing self-coupling strength of dark matter, particularly as the stellar structure transitions between dark core and dark halo configurations, depending on the total stellar mass. Nevertheless, we find that in the dark core structure, the fundamental ($f$) mode frequencies associated with dark matter exceed those of normal matter, at least when the central energy densities of both fluids are equal. Furthermore, we find that the $f$-mode frequencies associated with normal matter in dark core configurations adhere to a universal relation between the mass-scaled frequency and stellar compactness.
\end{abstract}

%
\maketitle


\section{Introduction}
\label{sec:I}

Neutron stars, formed as massive remnants of core-collapse supernovae, provide a unique natural setting to probe physics under extreme physical states. For instance, the density inside the star significantly exceeds the nuclear saturation density, and their gravitational and magnetic field strength can reach magnitudes far beyond those observed in our solar system \cite{ST83}. Since recreating such extreme-density environments on Earth is challenging due to nuclear saturation constraints, observations of neutron stars serve as a crucial means to explore the behavior of matter in extreme regimes. In practice, the discovery of massive neutron stars places strong constraints on the equation of state (EOS), effectively ruling out softer EOS models that cannot support the observed maximum masses~\cite{D10, A13, C20, F21, Romani22}. 

The detection of gravitational waves radiated from the binary neutron star merger event GW170817 provided constraints on the tidal deformability \cite{GW170817}, leading to an upper limit of 13.6 km for the radius of a $1.4M_\odot$ neutron star~\cite{Annala18}. The Neutron Star Interior Composition Explorer (NICER), which is a soft X-ray timing instrument aboard the International Space Station, has placed further constraints on the neutron star properties by precisely observing the pulsar light curve. This constraint arises because photons emitted from the neutron star's surface experience trajectory bending due to the strong gravitational field generated by the neutron star. As a result, NICER has placed constraints on the mass and radius of PSR J0030+0451~\cite{Riley19, Miller19} and PSR J0740+6620~\cite{Riley21, Miller21}. While these astronomical observations primarily probe neutron star properties at relatively high densities, terrestrial experiments remain essential for determining matter properties at lower densities (e.g., \cite{SNN22, SO22, SN23}).

In addition to constraints derived from gravitational waves and X-ray timing, the frequencies of stellar oscillations, if observed, offer a complementary approach to probing the internal structure of neutron stars. Since oscillation frequencies are highly sensitive to the properties of the stellar interior, their detection can provide a means to infer the internal composition (or information) through inverse modeling. This technique, known as asteroseismology (or gravitational wave asteroseismology when applied to gravitational wave data or signals) is analogous to Earth's seismology and the Sun’s helioseismology. Each oscillation mode carries distinct information about the star's physical state, allowing for the identification of underlying physics by matching observed frequencies with theoretical mode predictions. In practice, the crust EOS parameters and/or the stellar model have been constrained by associating the quasi-periodic oscillations observed in magnetars with the crustal torsional oscillations \cite{GNHL2011, SNIO2012, SIO2016, SKS23, Sotani24a}. It has been also suggested that future detections of gravitational waves from neutron stars may enable precise determinations of key properties such as mass, radius, compactness, tidal deformability, and the EOS (e.g.,~\cite{AK1996, AK1998, STM2001, SH2003, TL2005, SYMT2011, PA2012, DGKK2013, Sotani20b, Sotani21, SK21}). Similarly, the gravity ($g$-) modes, which arise from thermal gradients within the star, may also provide crucial insights into its thermal evolution history \cite{KHA15, SD22, SD24}. Furthermore, advancements in core-collapse supernova simulations have facilitated studies of gravitational waves emitted during the formation of protoneutron stars. These supernova-generated signals, when analyzed through asteroseismology, can shed light on the dynamic processes occurring in newly formed neutron stars, offering additional constraints on their structure and evolution (e.g.,~\cite{FMP2003, FKAO2015, ST2016, ST2020a, SKTK2017, MRBV2018, SKTK2019, TCPOF19, SS2019, ST2020, STT2021, SMT24}).

In addition to normal matter, dark matter (which constitutes $\sim 85\%$ of the matter content of the Universe~\cite{Plank_VI, Group22}) is expected to become an additional structural component for compact objects. The existence of dark matter is observationally well-established through the large-scale structures, galactic dynamics, and cosmic microwave background \cite{Davis85, Spergel03, Bertone05}. While its impact in the context of neutron stars remains uncertain, the intense gravitational potential of these stars may lead to the accumulation of dark matter within them, forming what are known as dark matter admixed neutron stars. The mechanism by which dark matter is captured by neutron stars is not yet fully understood, but the structural properties of these admixed stars are strongly influenced by the nature (or characteristics) of dark matter. This dependence suggests that observational signatures of neutron stars could potentially provide insights into the existence and characteristics of dark matter~\cite{LF10}.

Dark matter admixed neutron stars have recently gained significant attention, with numerous theoretical studies investigating their structural and dynamical properties under different dark matter models — such as self-interacting fermionic dark matter~\cite{Nelson19, Ivanytskyi20, SM24,  Rutherford24, KGS25}; asymmetric interacting or non-interacting bosonic dark matter~\cite{Ellis18, Fan12, KSST24, Konstantinou24}; dark matter interacting with nuclear matter via Higgs channel~\cite{PL17, DKP21, DKKP22, LLFD22, KS24}. Furthermore, the non-radial oscillations in these models have also been studied within single fluid formalism \cite{Das21, Flores24, Sirke24, Thakur24}. However, in cases where dark matter is only gravitationally coupled to normal matter without any other direct interactions, the oscillations must be treated as a two-fluid system, analogous to the Tolman-Oppenheimer-Volkoff (TOV) equations in the two-fluid formalism \cite{Goldman13, Sagun23}.

The two-fluid formalism has already been discussed in the context of neutron superfluidity \cite{Comer99,Comer02,Andersson02,Comer03,Comer04}, where the uncharged matter (neutron fluid) and charged matter (proton fluid) are considered as different constituents to discuss the possibility for additional mode excitation(s).
Even in the context of dark matter admixed neutron stars, the two-fluid formalism has been discussed for the tidal deformability and radial perturbations \cite{Leung11,Leung12,Leung22}.
In this study, to address this scenario, we derive a new set of perturbation equations under the Cowling approximation and compute the non-radial oscillation frequencies for dark matter admixed neutron stars, specifically adopting the self-interacting fermionic dark matter model.

This manuscript is organized as follows. In Sec. \ref{sec:NS_multi}, we present a brief derivation of the TOV equations for a multifluid system. Sec. \ref{sec:perturbation} introduces the perturbation equations formulated within the multifluid system. In Sec. \ref{sec:Application}, we apply these equations to dark matter admixed neutron stars, demonstrating the equilibrium models and corresponding oscillation modes. Finally, we summarize our findings and conclusions in Sec. \ref{sec:Conclusion}. Unless otherwise mentioned, we adopt geometric units with $c=G=1$, where $c$ and $G$ denote the speed of light and the gravitational constant, and use the metric signature $(-,+,+,+)$.

\section{Neutron star modes in multifluid formalism}
\label{sec:NS_multi}

We consider a static, spherically symmetric neutron star model, described by the following metric
\begin{equation}
  ds^2 = -e^{2\Phi}dt^2 + e^{2\Lambda}dr^2 + r^2 (d\theta^2 + \sin^2\theta d\phi^2), \label{eq:metric}
\end{equation}
where the metric functions, $\Phi$ and $\Lambda$ depend only on the radial coordinate, $r$, and $\Lambda$ is directly associated with the mass function, $m(r)$, which represent  the total gravitational mass enclosed within a radius $r$, through the relation $e^{-2\Lambda}=1-2m/r$. This equation ensures that the metric function accounts for the spacetime curvature due to the mass distribution inside the star. In the presence of dark matter that interacts only gravitationally with normal matter, the system can be modeled as a multifluid configuration. Each component (normal matter and dark matter) can be treated as an independent perfect fluid, given the absence of non-gravitational interactions. The energy density and pressure for each fluid component $x$ are denoted by $\varepsilon_{x}$ and $p_{x}$, respectively, and the total energy density and pressure of the system are given by: $\varepsilon_{\rm T} = \sum_x\varepsilon_{x}$ and $p_{\rm T} = \sum_x p_{x}$. Since each component behaves as a perfect fluid, the total energy-momentum tensor is expressed as
\begin{equation}
  T^{\mu\nu}_{\rm T} = (\varepsilon_{\rm T} + p_{\rm T})u^\mu u^\nu + p_{\rm T}g^{\mu\nu}, \label{eq:Tmunu0} 
\end{equation}
where $u^\mu$ denotes the four-velocity of the fluid elements. Owing to the decoupling between different fluids (except for the gravitational coupling), the total energy-momentum tensor, $T^{\mu\nu}_{\rm T}$, can be separated into the contributions from each fluid component, denoted by $T_{x}^{\mu\nu}$: 
\begin{gather}
  T^{\mu\nu}_{\rm T} = \sum_x T^{\mu\nu}_{x}, \label{eq:Tmunu1} \\
  T^{\mu\nu}_{x} \equiv (\varepsilon_{x} + p_{x})u_{x}^\mu u_{x}^\nu + p_{x}g^{\mu\nu}, \label{eq:Tmunux} 
\end{gather}
where $u^\mu_{x}$ denotes the four-velocity for the fluid $x$. We note that for a static stellar model, the equilibrium four-velocity satisfies:
\begin{equation} u^t = u^t_{x} = e^{-\Phi}, \quad u^i_{x} = 0 \quad \text{for} \quad i = r, \theta, \phi. \end{equation}
This ensures that in the static configuration, there is no motion of the fluid components relative to the coordinate system. From the energy-momentum conservation law, $\nabla_\mu T^{\mu\nu}_{\rm T}=0$ together with Eq. (\ref{eq:Tmunu1}), one can get the conservation law for each fluid, i.e., $\nabla_\mu T^{\mu\nu}_{x}=0$. Additionally, in equilibrium, the four-velocity satisfies $u^\mu_{\ ;\mu}=0$, ensuring stationarity.

From the Einstein equation, the mass $m(r)$  and metric $\Phi(r)$ function satisfies:
\begin{gather}
   m' = 4\pi r^2\varepsilon_{\rm T}, \label{eq:mr} \\
   \Phi' = \frac{4\pi r^3p_{\rm{T}} + m}{r(r-2m)}, \label{phir}
\end{gather}
where the prime denotes the ordinary differentiation with respect to $r$. Additionally, from the conservation of energy-momentum for each fluid component, $\nabla_\mu T_{x}^{\mu\nu}=0$, one can obtain:
\begin{align}
  p_x' = -\Phi' (\varepsilon_x + p_x). \label{eq:pr}
\end{align}
Combining Eqs. (\ref{phir}) and (\ref{eq:pr}), we can obtain the hydrostatic balance equation:
\begin{equation}
  p_x' = -\frac{(4\pi r^3p_{\rm T} + m)(\varepsilon_x + p_x)}{r(r-2m)}
\end{equation}
Since the total mass function can be decomposed into the contributions from each fluid, we write:
\begin{equation}
  m(r) = \sum_x m_x(r), 
\end{equation}
where $m_x(r)$ is the gravitational mass composed of the fluid $x$ inside the position $r$. Consequently, Eq. (\ref{eq:mr}) can also be expressed for each fluid component as
\begin{align}
   m_x' = 4\pi r^2\varepsilon_{x}. \label{eq:mxr}
\end{align}
The final form of the TOV equations for a multifluid system is:
\begin{align}
   m_x' &= 4\pi r^2\varepsilon_{x}, \label{eq:dm} \\
   p_x' &= -\frac{(4\pi r^3p_{\rm T} + m)(\varepsilon_x + p_x)}{r(r-2m)}, \label{eq:dp} \\
   \Phi' &= \frac{4\pi r^3p_{\rm T} + m}{r(r-2m)}, \label{dphi}
\end{align}
which retains the same form as in the standard two-fluid formalism~\cite{Goldman13, Sagun23}. These equations describe the equilibrium structure of a neutron star consisting of multiple independent fluids interacting solely through gravity.

\section{Perturbation equations in multifluid formalism}
\label{sec:perturbation}

As a first step in this study, we adopt the Cowling approximation, where the metric perturbations are neglected during the fluid oscillations. This approximation simplifies the analysis by focusing solely on fluid perturbations while assuming that the background spacetime remains unchanged. So, the perturbation equations for non-radial oscillations are derived from the linearized conservation law, i.e., $\nabla_\mu \left(\delta T^{\mu\nu}_{\rm T}\right)=0$, which, due to the decoupling between the different fluids, leads to the independent linear conservation equations for each fluid component, i.e., $\nabla_\mu \left(\delta T^{\mu\nu}_{x}\right)=0$. Thus, in the Cowling approximation, the oscillation modes of each fluid are independent of the others, at least at the perturbative level.
The Lagrangian displacements for the fluid $x$ are generally given by 
\begin{gather}
  \xi_{x}^i(t,r,\theta,\phi) = \left(e^{-\Lambda}W_x, -V_x\partial_\theta, -\frac{V_x}{\sin^2\theta}\partial_\phi\right)\frac{1}{r^2}Y_{\ell k}(\theta,\phi), \label{eq:xi}
\end{gather}
where $W_x$ and $V_x$ are functions of $t$ and radial coordinate $r$, and $Y_{\ell k}(\theta,\phi)$ represents the spherical harmonics with the azimuthal quantum number $\ell$ and magnetic quantum number $k$. Similarly, the perturbations of the energy density and pressure for each fluid component are expressed as:
\begin{gather}
  \delta\varepsilon_{x} = \delta\varepsilon_{x}(t,r)Y_{\ell m}(\theta,\phi), \\
  \delta p_{x} = \delta p_{x}(t,r)Y_{\ell m}(\theta,\phi).
\end{gather}
From Eq. (\ref{eq:xi}), the perturbed four-velocity $\delta u_{x}^i$ for the fluid $x$ is given by 
\begin{gather}
  \delta u_{x}^i = \left(e^{-\Lambda}\dot{W}_x, -\dot{V}_x\partial_\theta, -\frac{\dot{V}_x}{\sin^2\theta}\partial_\phi\right)\frac{1}{r^2}e^{-\Phi}Y_{\ell k}(\theta,\phi), \label{eq:ux}  
\end{gather}
where the dot denotes the partial derivative with respect to time $t$. Due to the Cowling approximation, the perturbation in $u_{x}^{t}$ is set to zero, i.e., $\delta u_{x}^t=0$.

To describe the perturbations in terms of thermodynamic variables, we introduce the adiabatic index $\gamma_{x}$ for each fluid component:
\begin{equation}
  \gamma_x \equiv \frac{\partial \ln p_x}{\partial \ln n_x}\bigg|_s, \label{eq:gamma_x}
\end{equation}
where $n_x$ is the number density of fluid $x$, and $s$ represents entropy. The Lagrangian perturbation of pressure can then be expressed as:
\begin{align}
  \delta p_x  &= \gamma_x p_x\frac{\Delta n_x}{n_x} - \frac{\partial p_x}{\partial r}\xi_x^r,  \label{eq:dpx}
\end{align}
where $\Delta n_x$ denotes its Lagrangian perturbation.
On the other hand, assuming an adiabatic process with no particle creation or annihilation, one can obtain the following equation from the 1st law in thermodynamics:
\begin{align}
   \Delta \varepsilon_x = (\varepsilon_x + p_x)\frac{\Delta n_x}{n_x},  \label{eq:dex}
\end{align}
where $\Delta\varepsilon_x$ is the Lagrangian perturbation of the energy density $\varepsilon_x$.
Combining Eqs. (\ref{eq:dpx}) and (\ref{eq:dex}), the perturbed pressure can be written as: 
\begin{align}
   \delta p_x =  c_{s,x}^2\delta \varepsilon_x + \frac{\gamma_x p_x {\cal A}_x}{r^2}e^{-\Lambda}W_x,  \label{eq:delta_p}
\end{align}
where the sound velocity, $c_{s,x}$, and ${\cal A}_x$ for fluid $x$ are given by 
\begin{align}
  c_{s,x}^2 &\equiv \frac{\partial p_x}{\partial \varepsilon_x}\bigg|_s = \frac{\gamma_x p_x}{\varepsilon_x + p_x}, \\
  {\cal A}_x(r) &= \frac{1}{\varepsilon_x + p_x}  \left(\varepsilon_x' - \frac{ p_x'}{c_{s,x}^2}\right).
\end{align}

In practice, the $t$-, $r$-, and $\theta$-components of the perturbed energy-momentum conservation law yield:
\begin{gather}
   \delta \varepsilon_x = -\frac{\varepsilon_x+p_x}{r^2}\left[e^{-\Lambda}W'_x
       + \ell(\ell+1)V_x\right]  - \frac{\varepsilon_x'}{r^2} e^{-\Lambda}W_x, \label{eq:delta_0} \\
   (\varepsilon_x+p_x)e^{-2\Phi+\Lambda}\frac{\ddot{W}_x}{r^2}
       + \delta p_x'  + \Phi' (\delta\varepsilon_x + \delta p_x)=0, \label{eq:delta_1} \\
   \delta p_x = (\varepsilon_x + p_x)e^{-2\Phi} \ddot{V}_x. \label{eq:delta_2}
\end{gather}
From Eqs. (\ref{eq:delta_p}) and (\ref{eq:delta_0}), one can obtain
\begin{align}
    \delta p_x =& -\frac{\varepsilon_x+p_x}{r^2}c_{s,x}^2\left[e^{-\Lambda}W'_x
       + \ell(\ell+1)V_x\right] - \frac{p_x'}{r^2}e^{-\Lambda}W_x, \label{eq:delta_p1}
\end{align}
while from Eqs. (\ref{eq:delta_2}) and (\ref{eq:delta_p1}) one can derive 
\begin{align}
    W'_x  = \frac{1}{c_{s,x}^2}\left[\Phi' W_x +\omega_x^2 r^2 e^{-2\Phi+\Lambda}V_x\right] - \ell(\ell+1)e^{\Lambda}V_x.  \label{eq:dW}
\end{align}
Moreover, combining Eq.~(\ref{eq:delta_1}) with  Eqs.~(\ref{eq:delta_0}), (\ref{eq:delta_p1}), and (\ref{eq:dW}),  we obtain the following equation for $V_{x}^{'}:$
\begin{align}
   V'_x = -e^{\Lambda}\frac{W_x}{r^2} + 2\Phi' V_x  
       - {\cal A}_x\left[\frac{\Phi'}{\omega_x^2r^2} e^{2\Phi-\Lambda}W_x + V_x\right], \label{eq:dV}
\end{align}
where we assume the time-dependence of perturbations of the form, $\exp(i\omega_x t)$
\footnote{In this study, we consider a system composed of fluid $x$, which does not directly interact with the other fluid except through gravity. Under the Cowling approximation, the oscillations of fluid $x$ can be excited independently of the motion of the other fluid and vice versa. However, when metric perturbations are included (i.e., beyond the Cowling approximation), the perturbations of fluid $x$ are gravitationally coupled with those of the other fluid, requiring the expansion in terms of $\exp(i\omega t)$ instead of $\exp(i\omega_x t)$. More specifically, in such a case the perturbations of fluid $x$ are coupled to the other fluid through metric perturbations.}.
To determine the eigenvalues $\omega_x$, one must solve the eigenvalue problem by integrating Eqs.~(\ref{eq:dW}) and (\ref{eq:dV}) under appropriate boundary conditions.

To ensure physically meaningful solutions, proper boundary conditions must be imposed at:
\begin{enumerate}
    \item The stellar center ($r = 0$) :

    From the perturbation equations, Eqs.~(\ref{eq:dW}) and (\ref{eq:dV}), the behavior of $W_{x}$ and $V_{x}$  near the center is given by:
        \begin{align} 
                W_x(r) &= -{\cal C}\ell r^{\ell+1} + {\cal O}(r^{\ell+3}), \nonumber \\
                V_x(r) &= {\cal C}r^{\ell} + {\cal O}(r^{\ell+2}), \nonumber
        \end{align} 
        where $\cal C$ is an arbitrary constant. Thus in the vicinity of the stellar center, this leads to the relation: \begin{equation} W_x(r) = -\ell r V_x(r). \end{equation}
    \item The fluid surface ($r = R_{x}$) : 

    The Lagrangian pressure perturbation must vanish at the fluid boundary, i.e., $\Delta p_x = 0$. Using the relation: \begin{align} \Delta p_x = \delta p_x + p_x' \xi^r = \delta p_x + p_x' e^{-\Lambda} \frac{W_x}{r^2}, \nonumber \end{align} along with Eq.~(\ref{eq:delta_p1}), this boundary condition simplifies to: \begin{equation} e^{-\Lambda} W_x' + \ell(\ell+1) V_x = 0. \label{eq:bc} \end{equation} By substituting Eq. (\ref{eq:dW}) into Eq. (\ref{eq:bc}), the boundary condition at $r = R_{x}$ can be further rewritten as: \begin{equation} \Phi' W_x + \omega_x^2 r^2 e^{-2\Phi+\Lambda} V_x = 0. \label{eq:bc2} \end{equation}
\end{enumerate} 
Thus, the eigenvalue problem for the oscillation frequencies $\omega_{x}$ is fully formulated. By solving Eqs.~(\ref{eq:dW}) and (\ref{eq:dV}) under the conditions specified above, one can determine the non-radial oscillation modes for two-fluid dark matter admixed neutron stars.
\section{Application}
\label{sec:Application}

In this study, we specifically investigate dark matter admixed neutron stars, where the dark matter interacts with normal (baryonic) matter solely through gravity, without any additional coupling. This scenario allows us to model the star as a two-fluid system, where each fluid component $x$ corresponds to either normal matter (NM) or dark matter (DM), following the formalism established in the previous section. With these models, we aim to explore the impact of dark matter on the non-radial oscillation frequencies of neutron stars. In particular, we examine how the presence of dark matter modifies the oscillation spectrum compared to standard neutron star models composed purely of normal matter. By quantifying these deviations, we aim to assess whether the presence of dark matter introduces distinct oscillation modes or shifts the fundamental frequencies in a measurable way.

\subsection{Dark Matter Model}
\label{sec:DMmodel}

Following previous studies~\cite{Nelson19, Ivanytskyi20, SM24, Rutherford24, KGS25}, we consider a self-interacting fermionic dark matter model with self-interactions mediated by a hidden gauge ${\cal V}^\mu$. In this scenario, dark matter is described by a Dirac fermion $\chi$, which interacts only within the dark sector and does not directly interact with normal (baryonic) matter through the Standard Model forces but only via gravitational coupling. The interaction within the dark sector is facilitated by a vector mediator, which influences the EOSs of dark matter inside neutron stars. The dynamics of fermionic dark matter and its self-interaction can be described by the following Lagrangian:
\begin{align}
    {\cal L}_{\rm DM} =& \,\,\bar\chi\,[ \gamma^\mu (i \partial_\mu - g_\chi {\cal V}_\mu ) - m_\chi] \,\chi \nonumber \\
    & -\frac{1}{2} m_{\rm v}^2 {\cal V}^\mu {\cal V}_\mu - \frac{1}{4} Z^{\mu \nu} Z_{\mu \nu} \ ,
    \label{Eq:LDS}
\end{align}
where $g_\chi$ denotes the dark sector gauge coupling constant; $m_\chi$ and  $m_{\rm v}$ denote the masses of the dark matter particle and the mediator. The field strength tensor of the mediator field is given by  $Z^{\mu \nu} = \partial^\mu {\cal V}^\nu-\partial^\nu {\cal V}^\mu$. Observational constraints on the self-interaction cross-section have been placed by studies of the Bullet Cluster (1E 0657-56) \cite{Randall08}, providing an upper limit on the interaction strength.

To describe the thermodynamic properties of dark matter in neutron stars, we use the mean-field approximation \cite{Rutherford24}, which allows the dark matter energy density and pressure to be expressed as:
\begin{align}
    \varepsilon_{\rm{DM}} &= \frac{2}{\left(2\pi\right)^{3}} \int^{k_{\chi}^{F}}_{0} \sqrt{k^{2}+m_{\chi}^{2}}\, d^{3}k \,+\,  \frac{1}{2} \left(\frac{g_{\chi}}{m_{\rm{v}}}\right)^{2} n_{\chi}^{2}, \label{eq:rhoDM} \\
    p_{\rm{DM}} &= \frac{2}{3\left(2\pi\right)^{3}} \int^{k_{\chi}^{F}}_{0} \frac{k^{2}}{\sqrt{k^{2}+m_{\chi}^{2}}}\, d^{3}k \,+\,  \frac{1}{2} \left(\frac{g_{\chi}}{m_{\rm{v}}}\right)^{2} n_{\chi}^{2}, \label{eq:pDM}
\end{align}
where $k_{\chi}^{\rm F}$ is the Fermi momentum of dark matter and $n_\chi$ represents the number density of dark matter particles, given by $n_{\chi} = \frac{2}{\left(2\pi\right)^{3}}\int^{k_{\chi}^{\rm F}}_{0} d^{3}k$. 
The first term in Eq.~(\ref{eq:rhoDM}) corresponds to the kinetic and rest-mass energy of dark matter particles, which follows the standard relativistic energy-momentum relation, and the second term accounts for the self-interaction energy mediated by ${\cal V}^\mu$, effectively introducing a repulsive interaction in the dark matter EOS. Similarly, in Eq.~(\ref{eq:pDM}), the first term represents the Fermi pressure, while the second term contributes to the pressure due to self-interactions. The repulsive force introduced by the vector mediator can influence the oscillation modes, as the presence of an additional fluid component alters the equilibrium structure and mode frequencies. By varying the coupling parameter $g_{\chi}/m_{\rm v}$, different dark matter EOS can be explored, allowing us to assess their impact on neutron star properties and oscillation spectra. 
We note that the case with $g_{\chi}/m_{\rm v}=0$ corresponds to a completely noninteracting dark matter model, i.e., the free fermi gas.

\subsection{Dark Matter admixed Neutron Star Models}
\label{sec:DMNSs}

To examine the oscillation frequencies of dark matter admixed neutron stars, we first construct the corresponding background stellar models. The modeling process requires selecting two key parameters from the dark matter sector, namely, the coupling ratio $g_\chi/m_v$ and the dark matter particle mass $m_\chi$, as well as, two central density parameters that characterize the neutron star, specifically the central energy densities of normal matter and dark matter, denoted as $\varepsilon_c^{\rm NM}$ and $\varepsilon_c^{\rm DM}$, respectively. Following the approach in Ref.~\cite{KGS25}, we set the dark matter particle mass to $m_\chi=5$ GeV, consistent with the estimation in Ref.~\cite{Zurek14} based on Plank observation, which suggests that the ratio of dark matter density to the baryon matter density in the Universe is approximately $5:1$ \cite{Plank}. For simplicity, in this study, we consider only the case where the central energy densities of normal matter and dark matter are equal, i.e. $\varepsilon_c^{\rm DM}=\varepsilon_c^{\rm NM}$, while models with different ratios of $\varepsilon_c^{\rm DM}/\varepsilon_c^{\rm NM}$ have been examined in Ref.~\cite{KGS25}. To construct the dark matter admixed neutron star models, a specific EOS for normal matter must be chosen. In this study, we adopt QMC-RMF4 EOS, derived in Ref.~\cite{QMC4}, within the framework of relativistic mean field theory. This EOS has been optimized for applications in neutron star astrophysics and successfully reproduces various astronomical observations. In particular, neutron star models built using the QMC-RMF4 EOS in the absence of dark matter are consistent with observational constraints on mass and radius (see Fig.\ref{fig:MRa}). The relevant EOS parameters and the expected maximum mass for neutron stars without dark matter are summarized in Table~\ref{tab:EOS0}. 

\begin{table}
\caption{EOS parameters, $K_0$, $L$, and $\eta\equiv(K_0L^2)^{1/3}$ \cite{SIOO14}, for the EOS adopted in this study together with the maximum mass of a spherically symmetric neutron star model without dark matter.} 
\label{tab:EOS0}
\begin{center}
\renewcommand{\arraystretch}{0.3} 
\setlength{\tabcolsep}{5pt} 
\begin{tabular}{ccccc}
\hline\hline
\vspace{0.25mm} \\ 
EOS & $K_0$ (MeV) & $L$ (MeV) & $\eta$ (MeV) & $M_{\rm max}/M_\odot$ \\
\vspace{0.25mm} \\ 
\hline \vspace{0.25mm} \\
QMC-RMF4  &  279  & 31.3  & 64.9  &  2.21  \\
\vspace{0.25mm} \\
\hline \hline
\end{tabular}
\end{center}
\end{table}

By solving the TOV equations for a two-fluid system, the neutron star's internal structure can be determined, including the boundaries of the normal matter and dark matter distributions. Specifically, the radius of the normal matter component $R_{\rm{NM}}$ is defined as the point where the pressure of normal matter $p_{\rm{NM}}$ vanishes, while the radius of the dark matter component $R_{\rm{NM}}$ is the point where the dark matter pressure $p_{\rm{DM}}$ vanishes. The mass contribution from each component is given by $M_{\rm NM} = m_{\rm NM}(R_{\rm NM})$ for the normal matter and $M_{\rm DM} = m_{\rm DM}(R_{\rm DM})$ for the dark matter. The total (gravitational) mass of the star, $M$, is then obtained as $M=M_{\rm NM} + M_{\rm DM}$, while the stellar radius, $R$, is defined as the outermost surface of the star, given by: $R={\rm max}(R_{\rm NM}, R_{\rm DM})$. Neutron stars with an admixture of dark matter exhibit two distinct structural configurations based on the relative distribution of normal and dark matter:
\begin{itemize}
    \item Dark core structure: $R_{\rm DM} < R_{\rm NM}$, where dark matter is concentrated within the core of the neutron star, forming a compact region at the center.
    \item Dark halo structure: $R_{\rm DM} > R_{\rm NM}$, where dark matter extends beyond the boundary of normal matter, forming an extended halo-like structure around the neutron star.
\end{itemize}
These structural differences have a significant impact on the star’s mass-radius relation, moment of inertia, and oscillation frequencies, providing potential observational signatures for the presence of dark matter within neutron stars.

\begin{figure}[tbp]
\begin{center}
\includegraphics[scale=0.6]{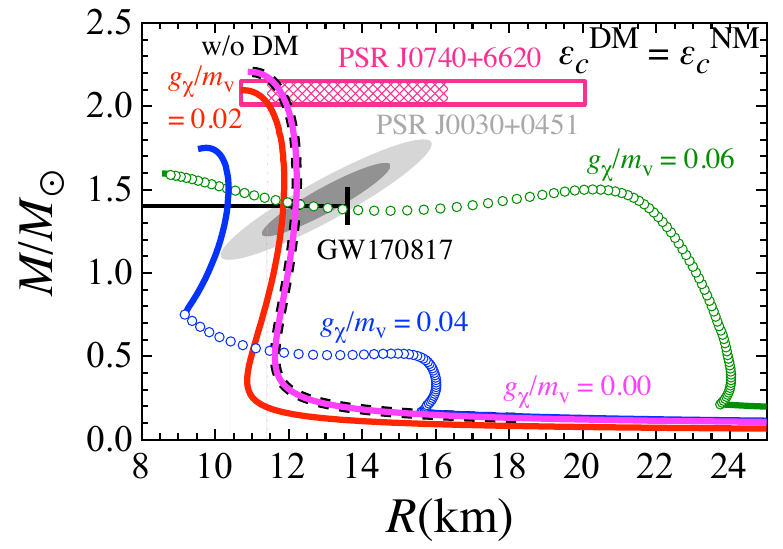} 
\end{center}
\caption{
Mass-radius relation for the dark matter admixed neutron star with equal central energy densities for normal and dark matter, i.e., $\varepsilon_c^{\rm DM}=\varepsilon_c^{\rm NM}$. The total mass is given by, $M=M_{\rm NM}+M_{\rm DM}$ while the stellar radius is defined as the outermost surface, i.e., $R={\rm max}(R_{\rm NM}, R_{\rm DM})$. Results are shown for dark matter coupling strengths $g_\chi/m_v=0.00$, 0.02, 0.04, and 0.06~MeV$^{-1}$, adopting the QMC-RMF4 EOS for normal matter. The solid lines correspond to the dark-core structures with $R_{\rm DM}<R_{\rm NM}$, whereas the open circles correspond to the dark-halo structures with $R_{\rm DM}>R_{\rm NM}$. For reference, the mass-radius relation for neutron star models without dark matter is also shown with the dotted line. Astronomical constraints from PSR J0740+6620, PSR J0030+0451, and GW170817 are also included (see also Fig.~\ref{fig:MRa} for the stellar models without dark matter).  
}
\label{fig:MR1}
\end{figure}

\begin{figure}[tbp]
\begin{center}
\includegraphics[scale=0.6]{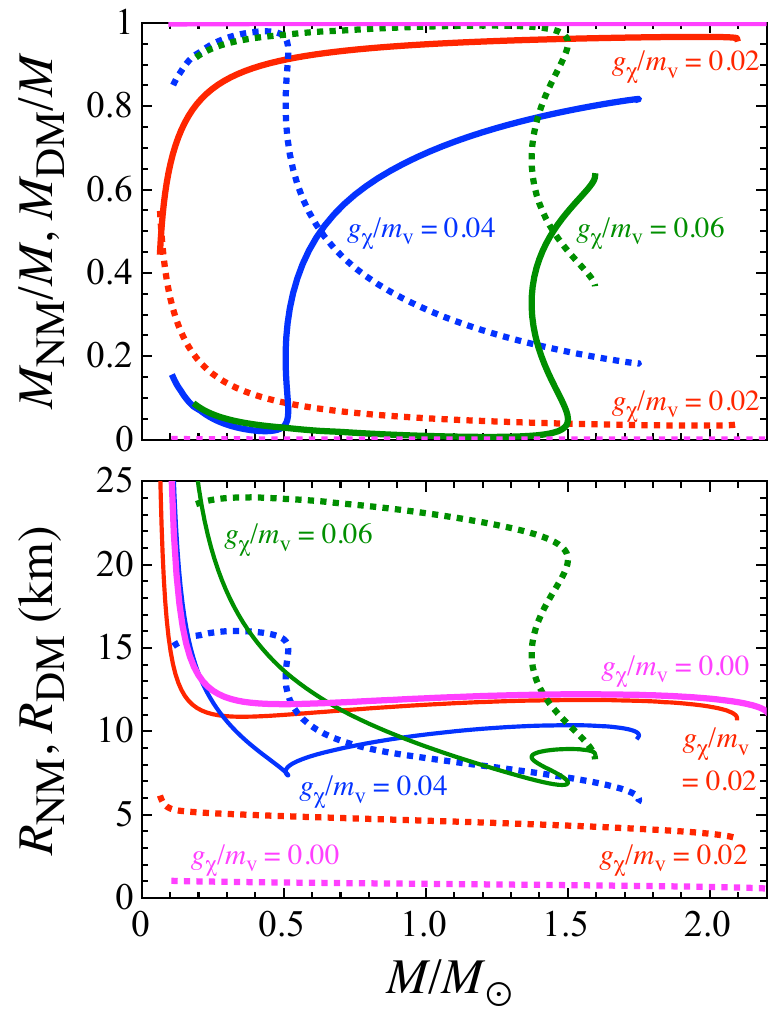} 
\end{center}
\caption{
The top panel presents the mass fractions of normal matter and dark matter, given by $M_{\rm NM}/M$ (solid lines) and $M_{\rm DM}/M$ (dotted lines) respectively, as functions of the total stellar mass. 
We note that, with $g_\chi/m_v=0$, $M_{\rm NM}/M\simeq 1$ while $M_{\rm DM}/M$ is only $10^{-4}-10^{-5}$. The bottom panel displays the corresponding radii of the normal matter and dark matter components, where $R_{\rm NM}$ (solid lines) denotes the radius of the normal matter distribution, and $R_{\rm DM}$ (dotted lines) represents the outer boundary of the dark matter component. 
}
\label{fig:RF_M}
\end{figure}

Figure~\ref{fig:MR1} presents the mass-radius ($M-R$) relation for the dark matter admixed stellar models with coupling ratios $g_\chi/m_v=0.00$, 0.02, 0.04, and 0.06~MeV$^{-1}$. For comparison, the dotted line represents the stellar model without dark matter. The solid lines correspond to neutron stars with a dark core structure ($R_{\rm DM} < R_{\rm NM}$), while the open circles denote models with a dark halo structure ($R_{\rm DM} > R_{\rm NM}$). For smaller values of the coupling ratio, such as $g_\chi/m_v=0.00$ and 0.02~MeV$^{-1}$, the impact of dark matter on the overall stellar structure is minimal, and the configuration remains predominantly a dark core. In fact, with $g_\chi/m_v=0.00$, the mass and radius for the dark matter admixed neutron star models are almost the same as those without dark matter, where the dark matter core is quite tiny (see also Fig.~\ref{fig:RF_M}). As the value of $g_\chi/m_v$ increases, the dark halo structure emerges, initially appearing at the left-bottom part of the $M-R$ diagram. This transition occurs because the self-interaction of dark matter increases, allowing dark matter to extend beyond the normal matter surface. Additionally, the maximum mass of the dark core configuration decreases, while that of the dark halo configuration increases. Finally, with a larger value of $g_\chi/m_v$, e.g., $g_\chi/m_v=0.06$~MeV$^{-1}$, the majority of the stellar structure exhibits dark halo configuration, indicating that dark matter dominates the overall structure and equilibrium of the star. However, the stellar models with $g_\chi/m_v=0.04$ and 0.06~MeV$^{-1}$, cannot satisfy the observational constraints imposed by PSR J0740+6620, one of the most massive neutron stars observed to date. Despite this limitation, these models are still valuable for understanding the qualitative effects of dark matter on non-radial oscillation frequencies. Therefore, in this study, we use these stellar models as the background configurations for our linear perturbation analysis. 

Figure~\ref{fig:RF_M} illustrates additional stellar properties of dark matter admixed neutron stars. The top panel displays the mass fractions of normal matter and dark matter, represented by $M_{\rm NM}/M$ (solid lines) and $M_{\rm DM}/M$ (dotted lines), respectively, as functions of the total stellar mass. Meanwhile, the bottom panel shows the corresponding radii of normal matter and dark matter distributions, where $R_{\rm NM}$ (solid lines) denotes the outer boundary of normal matter, and $R_{\rm DM}$ (dotted lines) represents the extent of dark matter. From this figure, it is evident that as the total mass increases, the contribution of dark matter becomes more significant in the dark halo configuration. This trend highlights the structural differences between dark core and dark halo neutron stars—while in dark core configurations, dark matter remains confined within the core, in dark halo structures, it extends beyond the normal matter surface, becoming the dominant component at larger masses. A more detailed analysis of how stellar properties depend on the dark matter interaction strength and central densities can be found in Ref.~\cite{KGS25}.

\subsection{Oscillation Frequencies}
\label{sec:frequency}

Using the neutron star models constructed in the previous section, we examine the non-radial oscillation frequencies by solving the eigenvalue problem associated with stellar perturbations. In this study, we simply assume that $c_{s,x}^2=dp_x/d\varepsilon_x$, which enables us to examine the excitation of both fundamental ($f$-) and the pressure ($p_i$-) modes. In particular, we focus on the quadrupolar ($\ell=2$) oscillations, restricting our analysis to the $f$-modes and the first overtone $p_1$-modes in this study. Due to the Cowling approximation, the oscillation modes of normal matter and dark matter are decoupled and can be expressed separately as $\omega^{\rm NM}$ and $\omega^{\rm DM}$, corresponding to the oscillations of normal matter and dark matter, respectively. Once these eigenvalues are determined, the corresponding oscillation frequencies are given by $f^{\rm NM}=\omega^{\rm NM}/2\pi$ and $f^{\rm DM}=\omega^{\rm DM}/2\pi$. A key consequence of the two-fluid formalism is the emergence of new oscillation modes associated with dark matter. The frequencies $f^{\rm{DM}}$ correspond to modes that arise exclusively due to the presence of dark matter within the neutron star. However, as we will see in the following analysis, the behavior of both $f^{\rm{NM}}$ and $f^{\rm{DM}}$ becomes increasingly complex in dark matter admixed neutron stars, especially as the interaction strength and distribution of dark matter influence the overall stellar dynamics.

\begin{figure}[tbp]
\begin{center}
\includegraphics[scale=0.6]{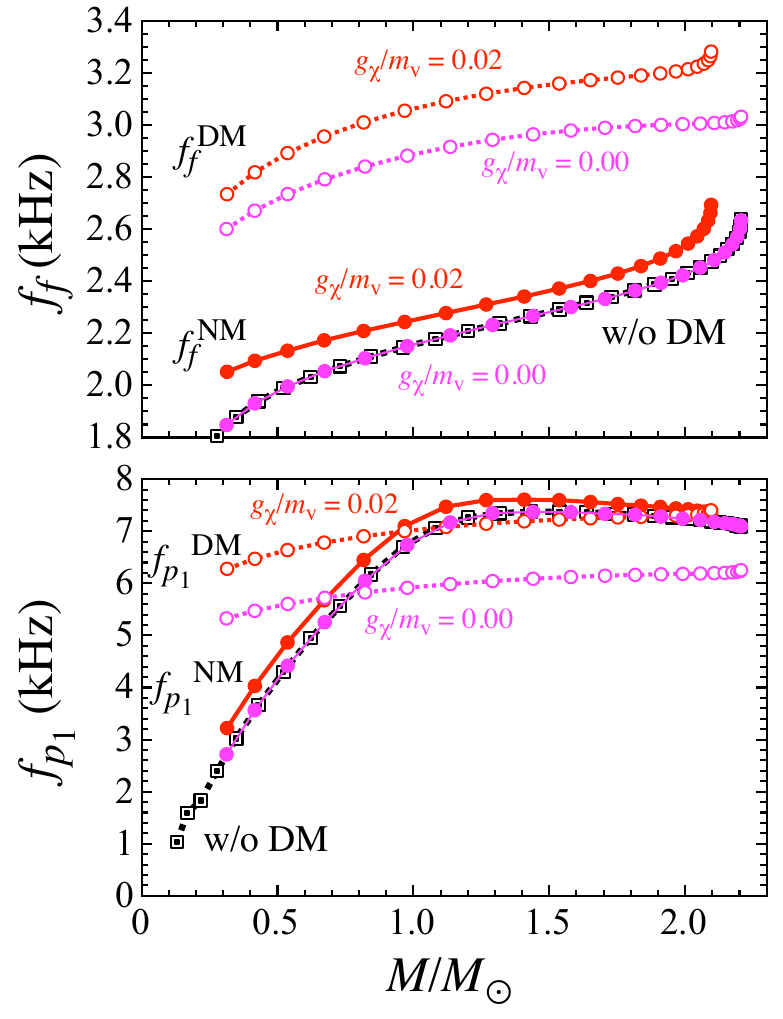} 
\end{center}
\caption{
The $f$- (top panel) and $p_1$-mode frequencies (bottom) panel for the dark matter admixed neutron star with $g_\chi/m_v=0.00$ and 0.02 MeV$^{-1}$ are shown as a function of the total mass. The solid lines with filled marks (dotted lines with open marks) correspond to the frequencies associated with normal matter (dark matter). For reference, the frequencies excited in the neutron star model without dark matter are also shown as dotted lines with double-square markers.
}
\label{fig:fBD1}
\end{figure}

\begin{figure}[tbp]
\begin{center}
\includegraphics[scale=0.6]{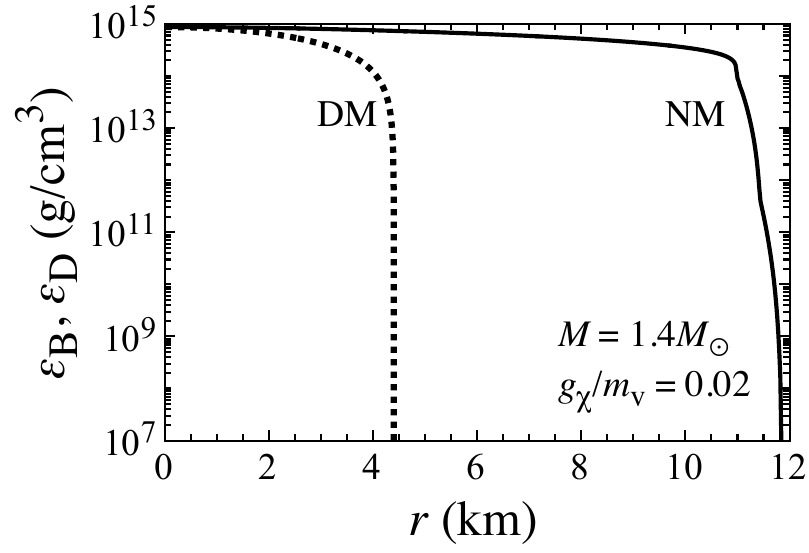} 
\end{center}
\caption{
Radial profile of the energy density for normal matter (solid line) and dark matter (dotted line) for the $1.4M_\odot$ stellar model with $g_\chi/m_v=0.02$ MeV$^{-1}$.
}
\label{fig:rhor}
\end{figure}

\begin{figure}[tbp]
\begin{center}
\includegraphics[scale=0.6]{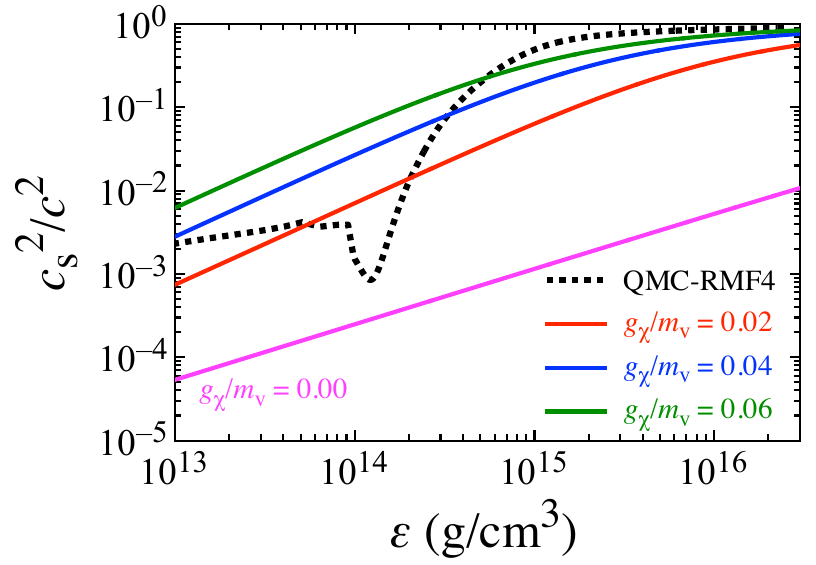} 
\end{center}
\caption{
Dependence of the sound velocity ($c_s^2$) on the energy density for the dark matter EOSs with $g_\chi/m_v=0.00$, 0.02, 0.04, and 0.06 MeV$^{-1}$, along with the QMC-RMF4 EOS for normal matter. 
}
\label{fig:cs2}
\end{figure}

Figure~\ref{fig:fBD1} presents the fundamental ($f$-mode) and first overtone ($p_1$-mode) oscillation frequencies, $f_f$ and $f_{p_1}$, as functions of the total stellar mass for neutron star models with $g_\chi/m_v=0.00$ and 0.02 MeV$^{-1}$. The solid lines with filled marks represent the oscillation frequencies associated with the normal matter ($f_f^{\rm NM}$ and $f_{p_1}^{\rm NM}$), while the dotted lines with the open marks correspond to those associated with the dark matter ($f_f^{\rm DM}$ and $f_{p_1}^{\rm DM}$). For comparison, the dotted lines with double squares indicate the frequencies obtained for stellar models without dark matter. As mentioned above, since the stellar models with $g_\chi/m_v=0.00$ and 0.02 MeV$^{-1}$ belong to the dark core configuration, where the structural deviation from the stellar models without dark matter is relatively small. Consequently, the oscillation frequencies, ($f_f^{\rm{NM}}$ and $f_{p_1}^{\rm{NM}}$) associated with normal matter follow a behavior similar to those of stellar models without dark matter. However, the frequencies associated with dark matter exhibit a more intricate dependence on the stellar mass. A notable feature is that the fundamental mode frequency of dark matter ($f_f^{\rm{DM}}$) is systematically larger than that of normal matter ($f_f^{\rm{NM}}$). This can be attributed to the fact that, in a dark core configuration, dark matter is highly confined to the inner regions of the star, leading to a steeper density gradient and a stronger restoring force for oscillations. This interpretation is supported by Fig.~\ref{fig:rhor}, which shows the radial profiles of energy density for normal matter (solid line) and dark matter (dotted line) for the $1.4M_\odot$ stellar model with $g_\chi/m_v=0.02$ MeV$^{-1}$. The plot clearly illustrates that dark matter is significantly more centrally concentrated than normal matter, reinforcing the idea that a more compact distribution leads to higher characteristic oscillation frequencies. 

The behavior of the pressure ($p_1$-) mode frequencies is more complex.
Even though the $p$-modes, which are kinds of acoustic oscillations, may be associated with the sound velocity (or stellar average density) as shown in Fig.~\ref{fig:cs2}, their frequency behavior becomes not so simple, depending on the position of the nodal point(s) in the wavefunctions. Nevertheless, from the bottom panel of Fig.~\ref{fig:fBD1}, we observe that 
\begin{itemize}
    \item For lower-mass neutron stars, the $p_{1}$-mode frequency associated with dark matter becomes higher than that with normal matter ($f_{p_{1}}^{\rm{DM}} > f_{p_{1}}^{\rm{NM}}$).
    \item For more massive neutron stars, 
    the $p_{1}$-mode frequency associated with dark matter weakly depends on the stellar mass, and eventually, it becomes lower than that with normal matter  ($f_{p_{1}}^{\rm{DM}} < f_{p_{1}}^{\rm{NM}}$).
\end{itemize}
This transition highlights the role of the EOS and sound velocity distribution in shaping the oscillation spectra of dark matter admixed neutron stars. The interplay between the density profile and sound speed gradient suggests that the observational detection of mode splitting between normal matter and dark matter oscillations could provide indirect evidence for the presence of a dark matter component within neutron stars.

\begin{figure}[tbp]
\begin{center}
\includegraphics[scale=0.6]{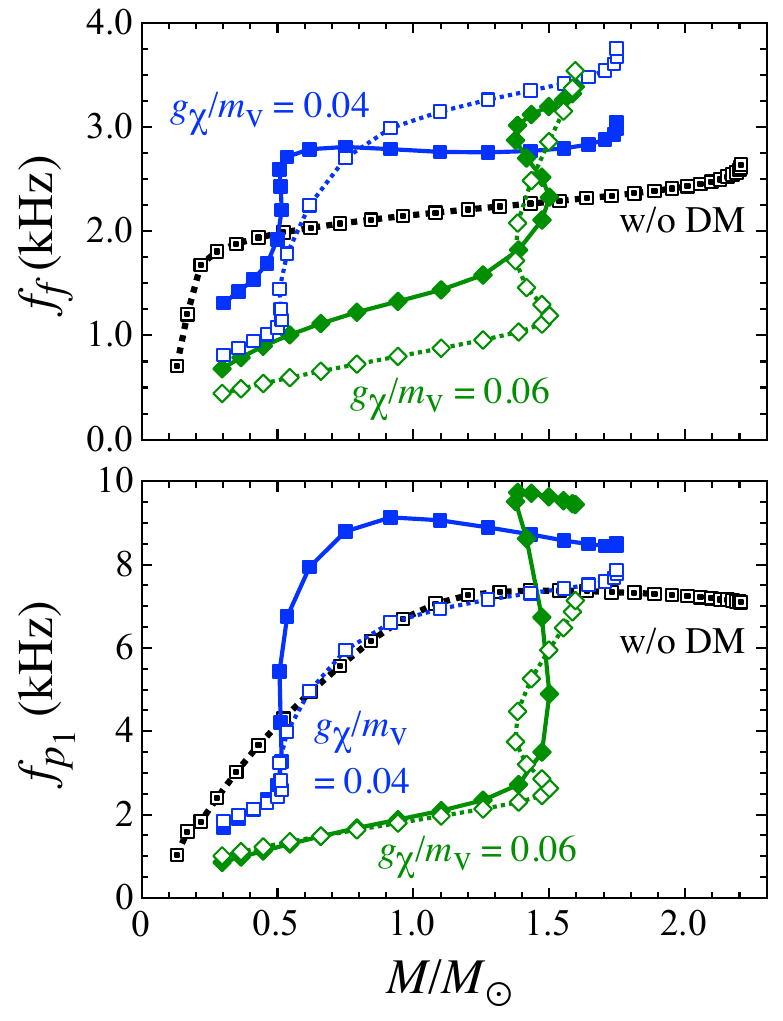} 
\end{center}
\caption{
Same as Fig.~\ref{fig:fBD1}, but for $g_\chi/m_v=0.04$ and 0.06 MeV$^{-1}$.
}
\label{fig:fBD2}
\end{figure}

The trends observed in the oscillation frequencies for the dark core configuration with $g_\chi/m_v = 0.00$ and 0.02 MeV$^{-1}$ are also present in stellar models with larger coupling ratios.  However, for higher values of $g_\chi/m_v$ the behavior becomes significantly more complex due to the transition between the dark core and dark halo structures, which depends on the total stellar mass. Figure~\ref{fig:fBD2} illustrates the $f$- and $p_1$-mode frequencies for neutron star models with $g_\chi/m_v=0.04$ and 0.06 MeV$^{-1}$, where the solid lines with filled marks represent the frequencies associated with normal matter, and the dotted lines with open marks correspond to those associated with dark matter. For comparison, the frequencies for neutron stars without dark matter are also plotted. 

As in the previous case, the $p_{1}$-mode frequency associated with dark matter ($f_{p_1}^{\rm DM}$) initially exceeds the corresponding mode in normal matter ($f_{p_1}^{\rm NM}$) at lower masses. Specifically, for $g_\chi/m_v=0.04$ MeV$^{-1}$, this occurs for masses below $M\lesssim 0.4M_\odot$, while for $g_\chi/m_v=0.06$ MeV$^{-1}$, the transition happens at $M\lesssim 0.7M_\odot$. This trend is likely to be associated with the dependence of the sound velocity on the energy density, as illustrated in Fig.~\ref{fig:cs2}. However, interpreting the results in the case of the $p_1$-mode is more challenging, since this mode exhibits a node in its eigenfunction, making it sensitive to the internal structure of the star across different density layers. In contrast, the behavior of the fundamental ($f$-mode) frequency is more directly linked to the stellar structure. As shown in Fig.~\ref{fig:rhor}, at least for models where $\varepsilon_c^{\rm DM}=\varepsilon_c^{\rm NM}$, the dark matter is more centrally concentrated than normal matter in the dark core configuration, whereas normal matter dominates the central region in the dark halo configuration. This structural difference leads to a distinct ordering of the fundamental mode frequencies: For dark core structure, where dark matter is more confined, the fundamental mode frequency of dark matter exceeds that of normal matter, i.e., $f_f^{\rm DM} > f_f^{\rm NM}$, whereas, for dark halo configurations, where the normal matter is more centrally concentrated, the opposite trend is observed, i.e. $f_f^{\rm DM} < f_f^{\rm NM}$. The transition point where $f_f^{\rm DM}$ and $f_f^{\rm NM}$ swap their relative magnitudes occurs at approximately $M\sim 0.8M_\odot$ for $g_\chi/m_v=0.04$ MeV$^{-1}$ and near the maximum mass $M\sim M_{\rm max}$ for $g_\chi/m_v=0.06$ MeV$^{-1}$. This shift in the ordering provides a clear structural signature of the transition between dark core and dark halo regimes and may serve as a key distinguishing feature in future astrophysical observations. 

As illustrated in Figs.~\ref{fig:fBD1} and~\ref{fig:fBD2}, the behavior of the eigenfrequencies excited in the dark matter admixed neutron stars exhibits significant complexity due to the interplay between normal and dark matter components. However, an alternative approach to understanding these oscillations more systematically is through the use of universal relations, which provide EOS-independent correlations between fundamental neutron star properties and their oscillation characteristics. Among the various universal relations associated with the $f$-mode frequencies, a particularly well-established one connects the mass-scaled $f$-mode frequency with the stellar compactness ($M/R$), as demonstrated in Ref.~\cite{TL2005}. The original relation has been derived without the Cowling approximation, i.e., including the metric perturbations. However, in this study, we examine the oscillation frequencies of the dark matter admixed neutron stars with the Cowling approximation, where metric perturbations are neglected. Therefore, it is essential to first establish the corresponding universal relation, adopting the Cowling approximation. By adopting the same functional form as in previous studies but recalibrating for the Cowling approximation, we obtain the following empirical relation:
\begin{equation}
   f_{f}M\ ({\rm kHz}/M_\odot) = -0.01932 + 2.115x + 1.731x^2 -0.5720x^3, \label{eq:fitting_noDM}
\end{equation}
where the dimensionless compactness parameter $x$ is defined as $x\equiv  \frac{M/R}{0.172}$. This normalization is chosen such that the compactness of a canonical neutron star model with $1.4M_\odot$ and a radius of 12 km corresponds to $x = 1$ (see Appendix~\ref{sec:appendix_1} for details on the derivation). 

\begin{figure*}[tbp]
\begin{center}
\includegraphics[scale=0.6]{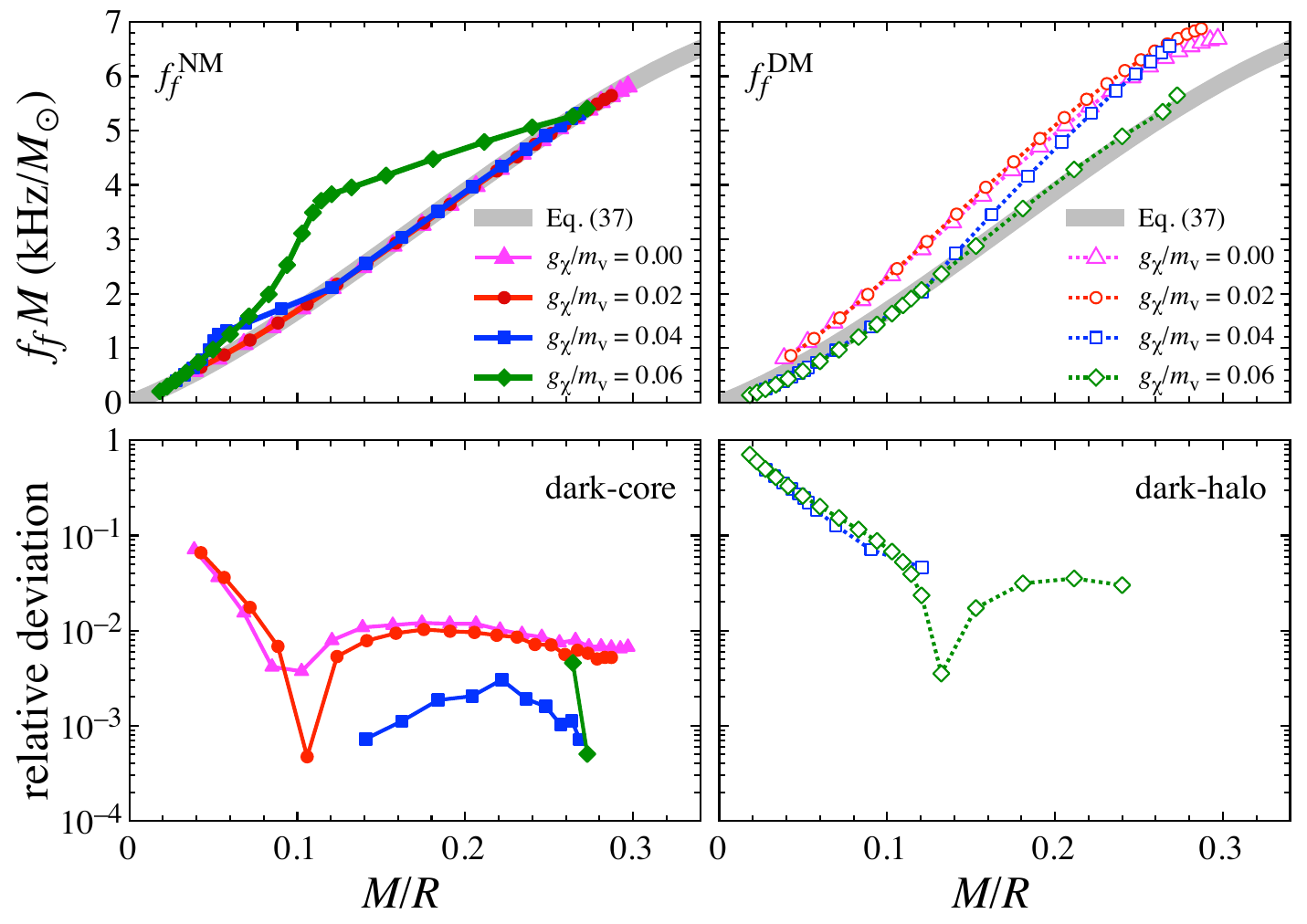} 
\end{center}
\caption{
In the top panel, the (total) mass-scaled $f$-mode frequencies are shown as a function of the stellar compactness defined with the total mass and radius, $R={\rm max}(R_{\rm NM}, R_{\rm DM})$, for the dark matter admixed neutron stars with $g_\chi/m_v=0.00$, 0.02, 0.04, and 0.06 MeV$^{-1}$, together with the universal relation for the neutron star models without dark matter given by Eq. (\ref{eq:fitting_noDM}). The left and right panels correspond to the $f$-mode frequencies associated with normal matter and dark matter, respectively. 
In the bottom panel, the relative deviation of the mass-scaled $f$-mode frequency from the fitting formula given by Eq. (\ref{eq:fitting_noDM}). The left-bottom panel corresponds to the $f$-mode frequencies associated with normal matter for the dark core model, while the right-bottom panel corresponds to those associated with dark matter for the dark halo model. 
}
\label{fig:ffM_MR}
\end{figure*}

To further explore the applicability of the mass-scaled universal relation in two-fluid dark matter admixed neutron stars, we examine how the fundamental ($f$-mode) frequencies behave as a function of stellar compactness.  Figure~\ref{fig:ffM_MR} presents the relation between the (total) mass-scaled $f$-mode frequencies as a function of the stellar compactness for neutron stars containing dark matter with coupling ratios $g_{\chi}/m_{v} = 0.00, 0.02, 0.04,$ and 0.06 MeV$^{-1}$. Here, the total stellar mass and stellar radius, defined as $R = \rm{max} (R_{\rm NM}, R_{\rm DM})$, are used to compute compactness. The left panel of Fig.~\ref{fig:ffM_MR} displays the results for oscillations associated with normal matter, while the right panel shows the frequencies corresponding to dark matter. From this analysis, we find that the fundamental mode frequency associated with normal matter ($f_{f}^{\rm NM}$) follows the established universal relation in the dark core configuration, despite the complex behavior observed in Figs.~\ref{fig:fBD1} and~\ref{fig:fBD2} (see the left-bottom panel in Fig.~\ref{fig:ffM_MR}). Similarly, for the dark halo configuration, the fundamental frequency associated with dark matter ($f_f^{\rm DM}$) also aligns reasonably well with the universal relation, though some deviations exist (see the right-bottom panel in Fig.~\ref{fig:ffM_MR}). This suggests that the mass-scaled universal relation, originally derived for single-fluid neutron stars, may still provide a reasonable approximation for oscillations in multi-fluid systems, depending on the stellar structure. However, given the limited number of stellar models considered in this study, a more systematic investigation is required to establish the extent to which the universal relation holds in dark matter admixed neutron stars. Future studies should explore wider parameter spaces, including variations in the central energy density ratio of dark matter-to-normal matter, different EOSs for normal matter, and non-Cowling approximations to gain deeper insights into the underlying physics governing neutron star oscillations.

\section{Conclusion}
\label{sec:Conclusion}

Dark matter admixed neutron stars represent a compelling astrophysical environment for investigating the properties of dark matter by comparing theoretical predictions with astronomical observations. In this study, we specifically explored self-interacting dark matter, where dark matter interacts with normal matter only through gravity, without any direct coupling via non-gravitational forces. To examine the non-radial oscillation frequencies excited in such a multi-fluid system, we derived the perturbation equations assuming the Cowling approximation. Using this framework, we examined a set of representative stellar models and determined the frequencies of both the fundamental ($f$-) and first pressure ($p_1$-) modes. A key outcome of this study is that the presence of dark matter leads to the excitation of additional oscillation modes associated with dark matter itself. In other words, alongside the standard $f$- and $p_1$-modes of normal matter ($f_{f}^{\rm NM}$ and $f_{p_{1}}^{\rm NM}$), the system also exhibits $f$- and $p_1$- modes associated with dark matter ($f_{f}^{\rm DM}$ and $f_{p_{1}}^{\rm DM}$). These additional modes emerge as a direct consequence of treating dark matter as an independent fluid component, and their properties depend on both the dark matter equation of state and its distribution within the star. 

For smaller/weaker dark matter self-interactions, the overall structure of dark matter admixed neutron star remains largely similar to that of a pure baryonic neutron star without dark matter admixture. In these cases, the star forms a dark core configuration, where dark matter is confined within the central region. Consequently, the oscillation frequencies associated with normal matter ($f_f^{\rm NM}$ and $f_{p_1}^{\rm NM}$) exhibit behavior that closely resembles that of ordinary neutron stars. However, as the dark matter coupling strength increases, the behavior of the eigenfrequencies becomes more complicated, because the stellar structure transitions from the dark core to dark halo configurations. Even in these complex scenarios, we identify a consistent trend: In dark core configurations, where dark matter remains more compactly distributed, the fundamental mode frequency associated with dark matter exceeds that of normal matter, i.e., $f_f^{\rm DM}>f_f^{\rm NM}$. Conversely, in dark halo configurations, where the normal matter is more centrally concentrated and dark matter extends beyond it, the relation is reversed, i.e. $f_f^{\rm DM}<f_f^{\rm NM}$. These trends hold at least when the central energy densities of dark matter and normal matter are equal. Additionally, our findings suggest that the universal relation between the mass-scaled $f$-mode frequencies and stellar compactness, originally established for standard (or pure baryonic) neutron stars, remains applicable to two-fluid dark matter admixed neutron stars. Specifically, in dark core configurations, the fundamental mode frequency associated with normal matter ($f_{f}^{\rm NM}$) adheres well to this relation. Meanwhile, for dark halo configurations, the fundamental mode frequency associated with dark matter ($f_{f}^{\rm NM}$) also appears to follow this relation, albeit with some deviations. While these initial results provide valuable insight into the oscillation properties of dark matter admixed neutron stars, a more systematic study is required. Future investigations should explore a wider range of stellar models, varying the dark matter interaction strength, mass fraction, and central energy densities, to fully characterize the role of dark matter in neutron star oscillations.


In this study, we adopt the Cowling approximation for simplicity. However, in a more realistic treatment—i.e., beyond the Cowling approximation—the oscillation modes of dark matter can couple to those of normal matter through metric perturbations, even when the dark matter interacts with normal matter only via gravity, as considered in Refs. \cite{Leung11,Leung12,Leung22}. In contrast, for dark matter models involving direct coupling with normal matter, one may extend the two-fluid formalism developed in the context of neutron star superfluidity \cite{Comer99,Comer02,Andersson02,Comer03,Comer04}, where entrainment effects can naturally be included. On the other hand, we note that our set of equations reduces to the relativistic multifluid equations originally derived in Ref.~\cite{Comer99} when the Cowling approximation is adopted and entrainment effects are neglected. Furthermore, although there is a systematic deviation between the mode frequencies obtained using the Cowling approximation and those computed within the full relativistic framework, a quantitative connection between the two has been established~\cite{Montefusco25}. Once the oscillation frequencies are computed without employing the Cowling approximation, a similar correspondence may be established even for dark matter admixed neutron stars.

\acknowledgments

This work is supported in part by Japan Society for the Promotion of Science (JSPS) KAKENHI Grant Numbers 
JP23K20848         
and JP24KF0090. 



\appendix
\section{Universal relation for the $f$-mode frequency without dark matter}   
\label{sec:appendix_1}

\begin{figure}[tbp]
\begin{center}
\includegraphics[scale=0.5]{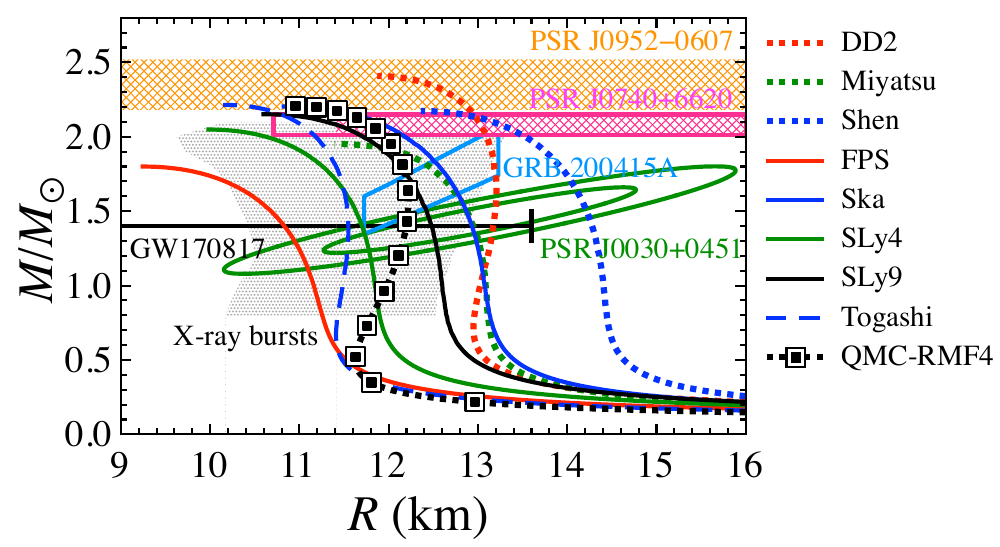} 
\end{center}
\caption{
Neutron star mass and radius profile without dark matter, using various EOSs listed in Table~\ref{tab:EOS}. For reference, constraints obtained from the astronomical observations are also shown, i.e., the mass of PSR J0952-0607 ($M=2.35\pm 0.17M_\odot$) \cite{Romani22} and PSR J0740+6620 ($M=2.08\pm 0.07M_\odot$) \cite{C20,F21}; the $1.4M_\odot$ radius constrained from the GW170817 event ($R_{1.4}\lesssim 13.6$ km) \cite{Annala18}; NICER observations for PSR J0030+0451 \cite{Riley19,Miller19} and PSR J0740+6620 \cite{Riley21,Miller21}; X-ray burst observations \cite{Steiner13}; and constraint from the magnetar high-frequency QPOs, GRB 200415A \cite{SKS23}. 
}
\label{fig:MRa}
\end{figure}

This study focuses on the oscillation frequencies of dark matter admixed neutron stars. To examine how the $f$-mode frequency is modified in the presence of dark matter, we derive the universal relations for the $f$-mode frequencies with the Cowling approximation, considering neutron star models without dark matter. These models are constructed using various EOSs listed in Table~\ref{tab:EOS} (see \cite{Sotani20b} for details on the EOSs adopted here). The oscillation frequencies for all EOSs, except for QMC-RMF4, are the results derived in Ref.~\cite{Sotani20b}, while the frequencies for QMC-RMF4 are newly derived in this study. It is worth noting that some of the EOSs considered here have already been ruled out by the astronomical observations as indicated in Fig.~\ref{fig:MRa}. However, we still include them in our analysis to explore the EOS dependence of the $f$-mode frequencies across a wide parameter range.

\begin{table}
\caption{EOS parameters, $K_0$, $L$, and $\eta\equiv(K_0L^2)^{1/3}$ \cite{SIOO14}, for the EOS adopted in Fig.~\ref{fig:univers_noDM} together with the maximum mass of spherically symmetric neutron stars without dark matter are listed. For reference, the data for QMC-RMF4 is also listed again.} 
\label{tab:EOS}
\begin {center}
\renewcommand{\arraystretch}{1.30} 
\setlength{\tabcolsep}{5pt}
\begin{tabular}{ccccc}
\hline\hline
EOS & $K_0$ (MeV) & $L$ (MeV) & $\eta$ (MeV) & $M_{\rm max}/M_\odot$    \\ 
\hline
DD2 & 243 & 55.0  & 90.2  & 2.41   \\
Miyatsu  &  274  & 77.1  & 118  & 1.95   \\
Shen & 281 & 111  &  151 & 2.17   \\
FPS  &  261  & 34.9  & 68.2  & 1.80   \\
SKa  &  263  & 74.6  & 114  & 2.22   \\
SLy4  &  230  & 45.9  & 78.5  & 2.05   \\
SLy9  &  230  & 54.9  &  88.4 & 2.16   \\
Togashi & 245 & 38.7 & 71.6 & 2.21   \\
\hline
QMC-RMF4  &  279  & 31.3  & 64.9  &  2.21  \\
\hline \hline
\end{tabular}
\end {center}
\end{table}

\begin{figure}[tbp]
\begin{center}
\includegraphics[scale=0.6]{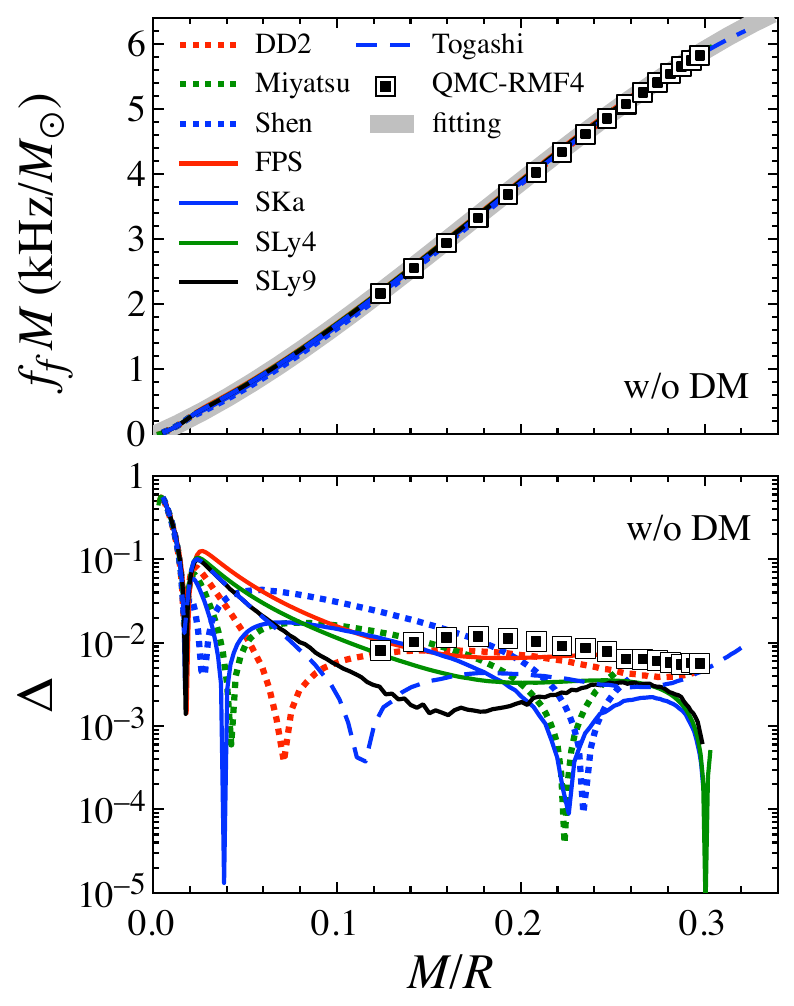} 
\end{center}
\caption{
The mass-scaled $f$-mode frequencies, $f_fM$, calculated with the Cowling approximation, are shown as a function of stellar compactness for various neutron star models without dark matter. The dotted lines denote the results with the EOSs based on the relativistic framework, the solid lines are those with the Skyrme-type effective interaction, and the dashed is those with Togashi EOS. The thick solid line is the fitting line given by Eq.~(\ref{eq:fitting_noDM}). In the bottom panel, we also show the relative deviation, $\Delta$, from the fitting formula, i.e., $\Delta=|f_f^{\rm eig}-f_f^{\rm fit}|/f_f^{\rm eig}$, where $f_f^{\rm eig}$ and $f_f^{\rm fit}$ respectively denote the $f$-mode frequencies determined from the eigenvalue problem and those estimated from the fitting formula.
}
\label{fig:univers_noDM}
\end{figure}

In the context of EOS-independent universal relations, it has been established that the mass-scaled $f$-mode frequencies can be expressed as a function of stellar compactness, $M/R$, largely independent of the specific EOS used \cite{TL2005}. In practice, as shown in Fig.~\ref{fig:univers_noDM}, the values of $f_f M$ calculated for various EOSs exhibit strong agreement, reinforcing the validity of this universal relation. We also derive the corresponding fitting formula given by Eq.~(\ref{eq:fitting_noDM}), with the expected values represented by the thick-solid line in the figure. To quantify the accuracy of this fitting, we define the relative deviation between the numerical eigenfrequency data and the fitted values as
\begin{equation}
  \Delta \equiv\frac{|f_f^{\rm eig}-f_f^{\rm fit}|}{f_f^{\rm eig}},
\end{equation}
where $f_f^{\rm eig}$ represents the $f$ mode frequency obtained from the eigenvalue problem, and $f_f^{\rm fit}$ is the frequency estimated using the fitting formula. From Fig.~\ref{fig:univers_noDM}, we find that the fitting performs well, achieving an accuracy of within $\sim 1\%$ for canonical neutron star models. However, for low-mass neutron stars, the fitting exhibits larger deviations. Although such low-mass models are not of primary astrophysical interest at present, future refinements could improve the accuracy of the fitting relation, potentially by incorporating higher-order terms in $M/R$.


\begin{thebibliography}{999}

\bibitem{ST83}
   S. L. Shapiro and S. A. Teukolsky, {\it Black Holes, White Dwarfs, and Neutron Stars: The Physics of Compact Objects}  (Wiley-Interscience, New York, 1983).

\bibitem{D10} 
   P. Demorest, T. Pennucci, S. Ransom, M. Roberts, and J. Hessels, Nature {\bf 467}, 1081 (2010).

\bibitem{A13} 
   J. Antoniadis {\it et al.}, Science {\bf 340}, 6131 (2013).

\bibitem{C20}    
   H. T. Cromartie {\it et al.}, Nature Astronomy {\bf 4}, 72 (2020).

\bibitem{F21}    
   E. Fonseca {\it et al.}, Astrophys. J. {\bf 915}, L12 (2021).

\bibitem{Romani22} 
   R. W. Romani, D. Kandel, A. V. Filippenko, T. G. Brink, and W. Zheng, Astrophys. J. {\bf 934}, L17 (2022).
 
\bibitem{GW170817}  
   B. P. Abbott {\it et al}. (LIGO Scientific and Virgo Collaborations), Phys. Rev. Lett. {\bf 119}, 161101 (2017).

\bibitem{Annala18}  
   E. Annala, T. Gorda, A. Kurkela, and A. Vuorinen, Phys. Rev. Lett. {\bf 120}, 172703 (2018).







\bibitem{Riley19} 
   T. E. Riley {\it et al.}, Astrophys. J.  {\bf 887}, L21 (2019).
  
\bibitem{Miller19} 
   M. C. Miller {\it et al.}, Astrophys. J.  {\bf 887}, L24 (2019).
   
\bibitem{Riley21} 
   T. E. Riley {\it et al.}, Astrophys. J.  {\bf 918}, L27 (2021).
   
\bibitem{Miller21} 
   M. C. Miller {\it et al.}, Astrophys. J.  {\bf 918}, L28 (2021).
   
\bibitem[Sotani, Nishimura, \& Naito (2022)]{SNN22}
   H. Sotani, N. Nishimura, and T. Naito, Prog. Theor. Exp. Phys. {\bf 2022}, 041D01 (2022).

\bibitem[Sotani \& Ota (2022)]{SO22}
   H. Sotani and S. Ota, Phys. Rev. D {\bf 106}, 103005 (2022).

\bibitem[Sotani \& Naito (2023)]{SN23}
   H. Sotani and T. Naito, Phys. Rev. C {\bf 107}, 035802 (2023).


\bibitem{GNHL2011}
   M. Gearheart, W. G. Newton, J. Hooker, and B. -A. Li, Mon. Not. R. Astron. Soc. {\bf 418}, 2343 (2011).
   
\bibitem{SNIO2012}
   H. Sotani, K. Nakazato, K. Iida, and K. Oyamatsu, Phys. Rev. Lett. {\bf 108}, 201101 (2012);
   Mon. Not. R. Astron. Soc. {\bf 428}, L21 (2013); {\bf 434}, 2060 (2013).

\bibitem{SIO2016}
   H. Sotani, K. Iida, and K. Oyamatsu, New Astron. {\bf 43}, 80 (2016);
   Mon. Not. R. Astron. Soc. {\bf 464}, 3101 (2017); {\bf 479}, 4735 (2018); 
   {\bf 489}, 3022 (2019).

\bibitem{SKS23}
   H. Sotani, K. D. Kokkotas, and N. Stergioulas, Astron. Astrophys. {\bf 676}, A65 (2023).

\bibitem{Sotani24a}  
   H. Sotani, Universe {\bf 10}, 231 (2024)

 
\bibitem{AK1996}
   N. Andersson and K. D. Kokkotas, Phys.\ Rev.\ Lett.\ {\bf 77}, 4134 (1996).

\bibitem{AK1998}
   N. Andersson and K. D. Kokkotas, Mon.\ Not.\ R. Astron.\ Soc.\ {\bf 299}, 1059 (1998).

\bibitem{STM2001}
   H. Sotani, K. Tominaga, and K. I. Maeda, Phys.\ Rev.\ D {\bf 65}, 024010 (2001).

\bibitem{SH2003}
   H. Sotani and T. Harada, Phys.\ Rev.\ D {\bf 68}, 024019 (2003);
   H. Sotani, K. Kohri, and T. Harada, {\it ibid}.\ {\bf 69}, 084008 (2004).

\bibitem{TL2005}
   L. K. Tsui and P. T. Leung, Mon.\ Not.\ R. Astron.\ Soc.\ {\bf 357}, 1029 (2005).

\bibitem{SYMT2011}
   H. Sotani, N. Yasutake, T. Maruyama, and T. Tatsumi, Phys.\ Rev.\ D {\bf 83} 024014 (2011).

\bibitem{PA2012}
   A. Passamonti and N. Andersson, Mon.\ Not.\ R. Astron.\ Soc.\ {\bf 419}, 638 (2012).

\bibitem{DGKK2013}
   D. D. Doneva, E. Gaertig, K. D. Kokkotas, and C. Kr\"{u}ger, Phys.\ Rev.\ D {\bf 88}, 044052 (2013).

\bibitem{Sotani20b}
   H. Sotani, Phys. Rev. D {\bf 102}, 063023 (2020); 103021 (2020).

\bibitem{Sotani21}
   H. Sotani, Phys. Rev. D {\bf 103}, 123015 (2021).

\bibitem{SK21}
   H. Sotani and B. Kumar, Phys. Rev. D {\bf 104}, 123002 (2021).

\bibitem{KHA15}
   C. J. Kr\"{u}ger, W. C. G. Ho, and N. Andersson, Phys.\ Rev.\ D {\bf 92}, 063009 (2015).
   
\bibitem{SD22}
   H. Sotani and A. Dohi, Phys. Rev. D {\bf 105}, 023007 (2022).

\bibitem{SD24}
   H. Sotani and A. Dohi, Phys. Rev. D {\bf 110}, 083036 (2024).

\bibitem{FMP2003}
   V. Ferrari, G. Miniutti, and J. A. Pons, Mon. Not. R. Astron. Soc. {\bf 342}, 629 (2003).

\bibitem{FKAO2015}
   J. Fuller, H. Klion, E. Abdikamalov, and C. D. Ott, Mon.\ Not.\ R. Astron.\ Soc.\ {\bf 450}, 414 (2015).

\bibitem{ST2016}
   H. Sotani and T. Takiwaki, Phys.\ Rev.\ D {\bf 94}, 044043 (2016); {\bf 102}, 023028 (2020).
   
\bibitem{ST2020a}
   H. Sotani and T. Takiwaki, Mon. Not. R. Astron. Soc. {\bf 498}, 3503 (2020).

\bibitem{SKTK2017}
   H. Sotani, T. Kuroda, T. Takiwaki, and K. Kotake, Phys.\ Rev.\ D {\bf 96}, 063005 (2017).

\bibitem{MRBV2018}
  V. Morozova, D. Radice, A. Burrows, and D. Vartanyan, Astrophys. J. {\bf 861}, 10 (2018).

\bibitem{SKTK2019}
   H. Sotani, T. Kuroda, T. Takiwaki, and K. Kotake, Phys.\ Rev.\ D {\bf 99}, 123024 (2019).

\bibitem{TCPOF19}
   A. Torres-Forn\'{e}, P. Cerd\'{a}-Dur\'{a}n, A. Passamonti, M. Obergaulinger, and J. A. Font, Mon. Not. R. Astron. Soc. {\bf 482}, 3967 (2019).

\bibitem{SS2019}
   H. Sotani and K. Sumiyoshi, Phys.\ Rev.\ D {\bf 100}, 083008 (2019); Mon. Not. R. Astron. Soc. {\bf 507}, 2766 (2021).

\bibitem{ST2020}
   H. Sotani and T. Takiwaki, Phys.\ Rev.\ D {\bf 102}, 063025 (2020).

\bibitem{STT2021}
   H. Sotani, T. Takiwaki, and H. Togashi, Phys.\ Rev.\ D {\bf 104}, 123009 (2021).

\bibitem{SMT24}
   H. Sotani, B. M\"{u}ller, and T. Takiwaki, Phys.\ Rev.\ D {\bf 109}, 123021 (2024).





\bibitem{Plank_VI}
   N. Aghanim et. al. (Planck Collaboration), Astron. Astrophys. {\bf 641}, A6 (2020).
   
\bibitem{Group22}   
   R. L.Workman et. al. (Particle Data Group), Prog. Theor. Exp. Phys. {\bf 2022}, 083C01 (2022).


\bibitem{Davis85}   
   M. Davis, G. Efstathiou, C. S. Frenk, and S. D. M. White, Astrophys. J. {\bf 292}, 371 (1985).
   
\bibitem{Spergel03} 
   D. N. Spergel et al., Astrophys. J. Suppl. {\bf 148}, 175 (2003).
   
\bibitem{Bertone05}  
   G. Bertone, D. Hooper, and J. Silk, Phys. Rep. {\bf 405}, 279 (2005).


\bibitem{LF10} 
   A. de Lavallaz and M. Fairbairn, Phys. Rev. D {\bf 81}, 123521 (2010).

\bibitem{Nelson19}
   A. E. Nelson, S. Reddy, and D. Zhou, JCAP {\bf 2019}, 012 (2019).

\bibitem{Ivanytskyi20}
   O. Ivanytskyi, V. Sagun, and I. Lopes, Phys. Rev. D {\bf 102}, 063028 (2020).
   
\bibitem{SM24}
   S. Shawqi and S. M. Morsink, Astrophys. J. {\bf 975}, 123 (2024).
   
\bibitem{Rutherford24}
   N. Rutherford, C. Prescod-Weinstein, and A. Watts, arXiv:2410.00140.

\bibitem{KGS25}  
   A. Kumar, S. Girmohanta, and H. Sotani, arXiv:2501.16829.


\bibitem{Ellis18}
   J. Ellis, G. H\"{u}tsi, K. Kannike, L. Marzola, M. Raidal, and V. Vaskonen, Phys. Rev. D {\bf 97}, 123007 (2018).
   
\bibitem{Fan12}
   Y.-z. Fan, R.-z. Yang, and J. Chang, arXiv:1204.2564 (2012).
   
\bibitem{KSST24}
   D. Rafiei Karkevandi, M. Shahrbaf, S. Shakeri, and S. Typel, Particles {\bf 7}, 201 (2024).
   
\bibitem{Konstantinou24}
   A. Konstantinou, Astrophys. J. {\bf 968}, 83 (2024).


\bibitem{PL17}
   G. Panotopoulos and I. Lopes, Phys. Rev. D 96, 083004 (2017).

\bibitem{DKKP22}
   H. C. Das, A. Kumar, B. Kumar, and S. K. Patra, Galaxies {\bf 10}, 14 (2022).

\bibitem{DKP21}
    H. C. Das, A. Kumar, and S. K. Patra, Phys. Rev. D {\bf 104}, 063028 (2021).
    
\bibitem{LLFD22}
   O. Louren\c{c}o, C. H. Lenzi, T. Frederico, and M. Dutra, Phys. Rev. D {\bf 106}, 043010 (2022).
   
\bibitem{KS24}
    A. Kumar and H. Sotani, Phys. Rev. D {\bf 110}, 063001 (2024).


\bibitem{Das21} 
   H. C. Das, A. Kumar, S. K. Biswal, and S. K. Patra, Phys. Rev. D {\bf 104}, 123006 (2021).

\bibitem{Flores24}  
   C. V. Flores, C. H. Lenzi, M. Dutra, O. Louren\c{c}o, J. D. V. Arba\~{n}il, Phys. Rev. D {\bf 109}, 083021 (2024).

\bibitem{Sirke24} 
   S. Shirke, B. K. Pradhan, D. Chatterjee, L. Sagunski, J. Schaffner-Bielich, PRD {\bf 110}, 063025 (2024).
 
\bibitem{Thakur24}
   P. Thakur, A. Kumar, V. B. Thapa, V. Parmar, and M. Sinha, J. Cosmol. Astropart. Phys. {\bf 12}, 042 (2024).

\bibitem{Goldman13}     
   I. Goldman, R. Mohapatra, S. Nussinov, D. Rosenbaum, and V. Teplitz, Phys. Lett. B {\bf 725}, 200 (2013).

\bibitem{Sagun23}     
   V. Sagun, E. Giangrandi, T. Dietrich, O. Ivanytskyi, R. Negreiros, and C. Provid\^{e}ncia, Astrophys. J. {\bf 958}, 49 (2023).




\bibitem{Comer99} 
   G. Comer, D. Langlois, and L. M. Lin, Phys. Rev. D 60, 104025 (1999).
   
\bibitem{Comer02} 
   G. L. Comer, Found. Phys. 32, 1903 (2002).
   
\bibitem{Andersson02} 
   N. Andersson, G. L. Comer, and D. Langlois, Phys. Rev. D 66, 104002 (2002).

\bibitem{Comer03} 
   G. L. Comer and R. Joynt, Phys. Rev. D 68, 023002 (2003).

\bibitem{Comer04} 
   G. Comer, Phys. Rev. D 69, 123009 (2004).
   

\bibitem{Leung11} 
   S.-C. Leung, M.-C. Chu, and L.-M. Lin, Phys. Rev. D 84, 107301 (2011).
   
\bibitem{Leung12} 
   S.-C. Leung, M.-C. Chu, and L.-M. Lin, Phys. Rev. D 85, 103528 (2012).
   
\bibitem{Leung22} 
   K.-L. Leung, M.-C. Chu, and L.-M. Lin, Phys.Rev.D 105, 123010 (2022).

\bibitem{Randall08}   
   S. W. Randall, M. Markevitch, D. Clowe, A. H. Gonzalez, and M. Bradac, Astrophys. J. {\bf 679}, 1173 (2008).




   
\bibitem{SIOO14} 
   H. Sotani, K. Iida, K. Oyamatsu, and A. Ohnishi, Prog. Theor. Exp. Phys. {\bf 2014}, 051E01 (2014).
  



   

\bibitem{Zurek14}  
   K. M. Zurek, Phys. Rept. 537, 91 (2014).

\bibitem{Plank}  
   P. Ade et al. (Planck Collaboration), Astron. Astrophys. {\bf 571}, A1 (2014).


\bibitem{QMC4}  
   M. G. Alford, L. Brodie, A. Haber, and I. Tews, Phys. Rev. C {\bf 106}, 055804 (2022).

\bibitem{Steiner13} %
   A. W. Steiner, J. M. Lattimer, and E. F. Brown, Astrophys. J. 765, L5 (2013).



\bibitem{Montefusco25} %
   G. Montefusco, M. Antonelli, and F. Gulminelli, Astron. Astrophys {\bf 694}, A150 (2025).

\end{thebibliography}

\end{document}